\documentclass[twocolumn,amsmath,amssymb,superscriptaddress,showkeys]{revtex4}
\usepackage[latin2]{inputenc}
\usepackage{graphicx}
\usepackage{dcolumn}
\usepackage{epstopdf}
\usepackage{amssymb,amsfonts,amsmath}

\def\SP{\hspace{4mm}}

\begin{document}

%%%%%%%%%%%%%%%%%%%%%%%%%%%%%%

\title{Kondo conductance across the smallest spin 1/2 
radical molecule}

\author{Ryan Requist}
\affiliation{SISSA, Via Bonomea 265, Trieste 34136, Italy}

\author{Silvio Modesti}
\affiliation{Physics Department, University of Trieste, Via Valerio 2, Trieste 34127, Italy}

\author{Pier Paolo Baruselli}
\affiliation{SISSA, Via Bonomea 265, Trieste 34136, Italy}
\affiliation{Institute for Theoretical Physics, TU Dresden, 01069 Dresden, Germany}
\affiliation{CNR-IOM Democritos, Via Bonomea 265, Trieste 34136, Italy}

\author{Alexander Smogunov}
\affiliation{CEA, IRAMIS, SPCSI, F-91191 Gif-sur-Yvette Cedex, France}

\author{Michele Fabrizio}
\affiliation{SISSA, Via Bonomea 265, Trieste 34136, Italy}
\affiliation{CNR-IOM Democritos, Via Bonomea 265, Trieste 34136, Italy}

\author{Erio Tosatti}
\affiliation{SISSA, Via Bonomea 265, Trieste 34136, Italy}
\affiliation{CNR-IOM Democritos, Via Bonomea 265, Trieste 34136, Italy}
\affiliation{ICTP, Strada Costiera 11, Trieste 34151, Italy}

%%%%%%%%%%%%%%%%%%%%%%%%%%%%%%

\begin{abstract} 

Molecular contacts are generally poorly conducting because their energy levels tend to lie 
far from the Fermi energy of the metal contact, necessitating undesirably large gate and 
bias voltages in molecular electronics applications.  Molecular radicals are an exception 
because their partly filled orbitals undergo Kondo screening, opening the way to electron 
passage even at zero bias.  While that phenomenon has been experimentally demonstrated 
for several complex organic radicals, quantitative theoretical predictions have not been 
attempted so far.  It is therefore an open question whether and to what extent an 
ab initio-based theory is able to make accurate predictions for Kondo temperatures and 
conductance lineshapes. Choosing nitric oxide NO as a simple and exemplary spin 1/2 
molecular radical, we present calculations based on a combination of density functional
theory and numerical renormalization group (DFT+NRG) predicting a zero bias spectral 
anomaly with a Kondo temperature of 15 K for NO/Au(111).  A scanning tunneling 
spectroscopy study is subsequently carried out to verify the prediction, and a striking zero 
bias Kondo anomaly is confirmed, still quite visible at liquid nitrogen temperatures.  
Comparison shows that the experimental Kondo temperature of about 43 K is larger than the 
theoretical one, while the inverted Fano lineshape implies a strong source of interference not 
included in the model.  These discrepancies are not a surprise, providing in fact an instructive 
measure of the approximations used in the modeling, which supports and qualifies the 
viability of the DFT+NRG approach to the prediction of conductance anomalies in larger 
molecular radicals.

\end{abstract}


\maketitle

Electron transport through molecules adsorbed on metallic surfaces or 
suspended between metal leads is the basic ingredient of molecular electronics 
\cite{aviram1974,reed1997,joachim2000}.  Because the highest occupied and 
lowest unoccupied molecular orbitals rarely lie close to the Fermi energy, electrons 
must generally tunnel through the molecule, making the zero bias conductance much 
smaller than $G_0 = 2 e^2/h$, the conductance quantum, whenever gating is not easily 
achieved, as is the case in mechanical break junctions and STM. That problem does not 
persist for molecular {\it radicals}, where one or more molecular orbitals are singly 
occupied, generally resulting in a net spin.  When brought into contact with a metal, the 
radical's spin is Kondo screened \cite{hewson1993}, leading to a zero bias conductance 
which may be of order $G_0$ with a Fano-like anomaly below the Kondo temperature 
$T_K$ \cite{madhavan1998,li1998} and no need for gating.   One reason for practical 
interest in such anomalies is the sensitivity of the conductance to external control 
parameters such as magnetic fields and mechanical strain \cite{parks2010}.  Several 
molecular contacts have been studied \cite{scott2010}, involving both adsorbed 
\cite{wahl2005,zhao2005,iancu2006,jiang2011,mugarza2011,muellegger2012,
muellegger2013,zhang2013} 
and contact-bridging \cite{park2002,parks2007,osorio2007,parks2010} molecules.  None 
so far involved a radical molecule that is both simple and spin 1/2, and the Kondo 
anomalies were not predicted from first principles, ahead of experiment.  Both the intrinsic 
complexity of the contact between a large molecular radical and a metal, and the 
unavailability of quantitatively tested {\it ab initio} electronic structure based approaches 
to Kondo conductance have so far restricted the theoretical work to the role of {\it a 
posteriori} support of STM and break junction zero bias anomaly data.  It is therefore 
important to achieve a first principles predictive capability of Kondo conductance anomalies 
across molecular radicals and ascertain its reliability.  To that end, we put to work a 
DFT+NRG method devised and implemented earlier in our group 
\cite{lucignano2009,baruselli2012a,baruselli2012b} but not yet verified by experiment. 
While other strategies with similar goals have also been proposed in Refs. 
\cite{costi2009,jacob2009,diasdasilva2009,surer2012}, we can now demonstrate the 
predictive power as well as the limitations of our procedure in a specific test case where 
direct comparison with experiment is possible.

Seeking a molecular radical Kondo system of the utmost simplicity and stimulated by the 
experimentally observed Kondo screening of the $S=1$ radical O$_2$ \cite{jiang2011}, 
we singled out nitric oxide, NO, as the smallest and simplest $S=1/2$ radical molecule 
which might display a Kondo conductance anomaly in an adsorbed state.  We have 
therefore applied our theoretical program to STM conductance of the NO radical adsorbed 
on the Au(111) surface.  NO is an interesting test case, in the past only qualitatively 
considered as a prototype of radical-surface Kondo interactions 
\cite{yoshimori1995,perezjigato1999}, while recently it has been used instead 
to quench the spin state of larger molecular radicals such as porphyrins and 
phthalocyanines \cite{waeckerlin2010}.

\begin{figure}[t!]
\centering
\includegraphics[width=0.75\columnwidth]{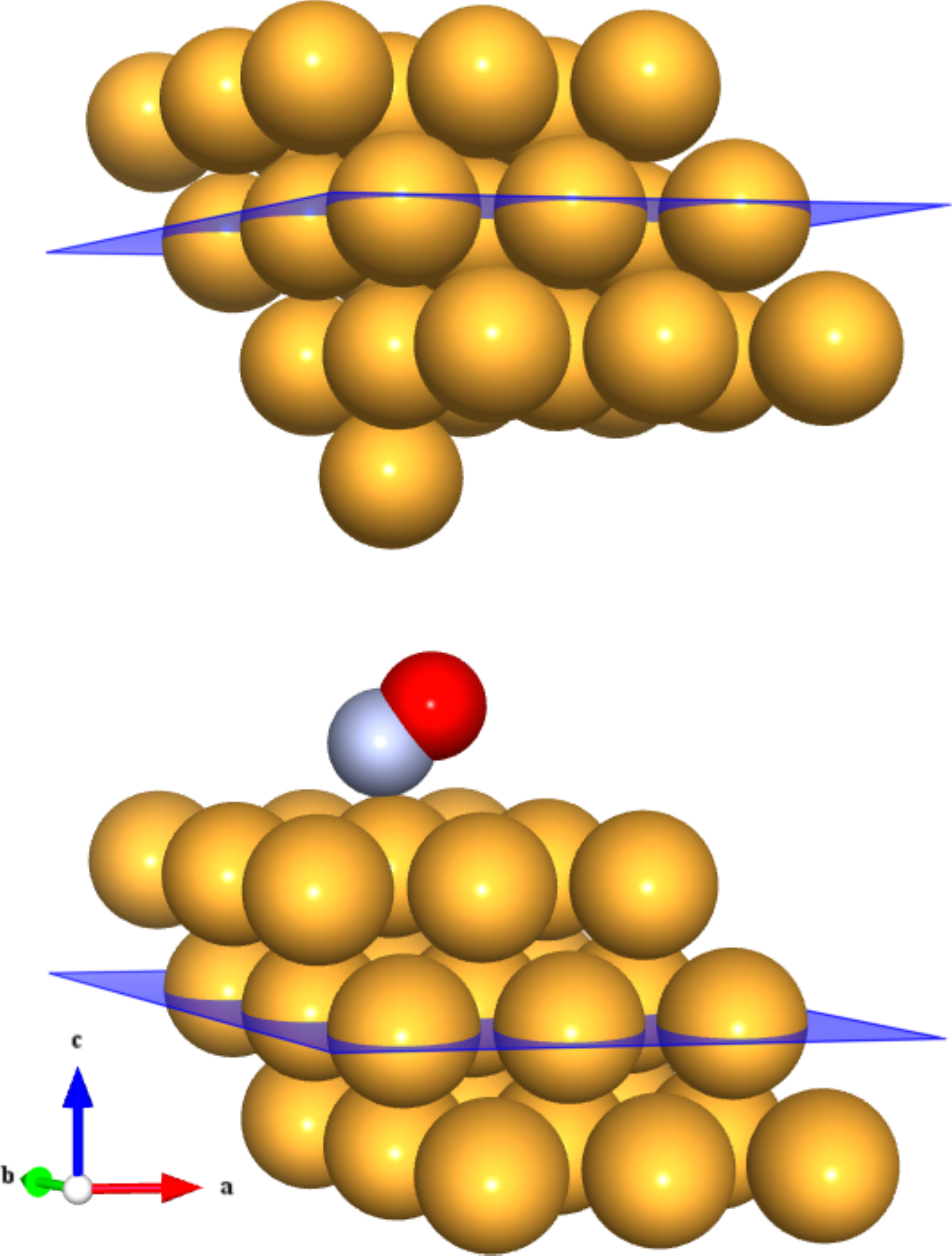}
\caption{\label{fig:scat} Model STM geometry for NO/Au(111).  The scattering region 
shown contains one NO molecule and 28 Au atoms; it satisfies periodic boundary conditions 
in the transverse direction and is bounded by two planes in the $c$ direction, where the 
potential is smoothly matched to the potential of bulk gold.  NO is shown in the optimal 
on-top adsorption geometry, where the molecule is bent by about 60 degrees.  Figure 
produced using VESTA \cite{vesta}.}
\end{figure}

Starting with DFT calculations of an idealized Au(111)/ NO/STM tip geometry as shown in 
Fig.~\ref{fig:scat}, we determined the adsorption geometry and the persistence of a 
nonzero spin state carried by the 2$\pi^*$ orbitals of NO.  
A two-orbital Anderson model 
describing their hybridization with the gold surface Fermi sea was constructed and its 
parameters adjusted to quantitatively reproduce the DFT results. The NRG solution of the 
Anderson model (\textit{NRG Results}) determined which of the two $2\pi^*$ orbitals, 
namely the one with odd symmetry, is eventually screened at the Kondo fixed point. A 
spectral function peak and a zero bias conductance anomaly were predicted with a Kondo 
temperature of about \mbox{15 K} (half width at half maximum (HWHM) of 5 meV) \footnote{Adopting Wilson's definition, the Kondo temperature $T_{K,W}$ is inferred from $\Delta_{\rm HWHM} \approx 4.6 k_B T_{K,W}$ \cite{zitko2009}, where the HWHM is extracted from the linewidth $\Gamma_K$ obtained from a Frota lineshape fit \cite{frota1992}.}.  To check these predictions we carried out STM/STS experiments (\textit{STM/STS Results}) for NO/Au(111).  A clear Kondo anomaly was found with an experimental Kondo temperature of about 43 K (HWHM$=$16 meV).  The observed conductance lineshape near zero bias, a sharp dip, is also quite close but exactly complementary to the calculated Kondo spectral peak.  The broad agreement with experiment and especially the discrepancies represent an important critical validation, pointing out physical variables to be included in future calculations.

\bigskip
\noindent \textbf{First Principles Calculations and Modeling}

\noindent The initial step is a DFT calculation (\textit{Materials and Methods} and \textit{SI Text}) of NO 
adsorbed on an unreconstructed Au(111) surface, adopted here as an approximation to the 
face-centered cubic (fcc) domains of the ``herringbone'' reconstructed Au(111) surface
\cite{barth1990}.  Our calculations show that the only stable NO adsorption geometry is 
on-top, with N pointing down toward a Au surface atom and, remarkably, a tilt angle of 
nearly 60 degrees as illustrated in Fig.~\ref{fig:scat}.  NO also binds (with lower binding 
energy) at Au(111) bridge and hollow sites.  However, in the low coverage limit these 
adsorption states are only stable when NO is constrained to be upright, for otherwise there 
is no barrier against migration to the optimal tilted on-top adsorption state. The calculated 
on-top adsorption energy 320 meV is weak and compares well with an estimated 400 meV 
from temperature programmed desorption spectroscopy \cite{mcclure2004}. 

Consistent with that weak adsorption state, spin-polarized DFT shows the persistence of a 
magnetic moment close to one Bohr magneton, meaning that adsorbed NO retains its 
radical character as in vacuum.  Both $2\pi^*$ molecular orbital resonances, their 
degeneracy lifted in the tilted on-top adsorption state, cross the gold Fermi level 
(Fig.~\ref{fig:dos-top}).  
\begin{figure}[t!]
\centering
\includegraphics[width=\columnwidth]{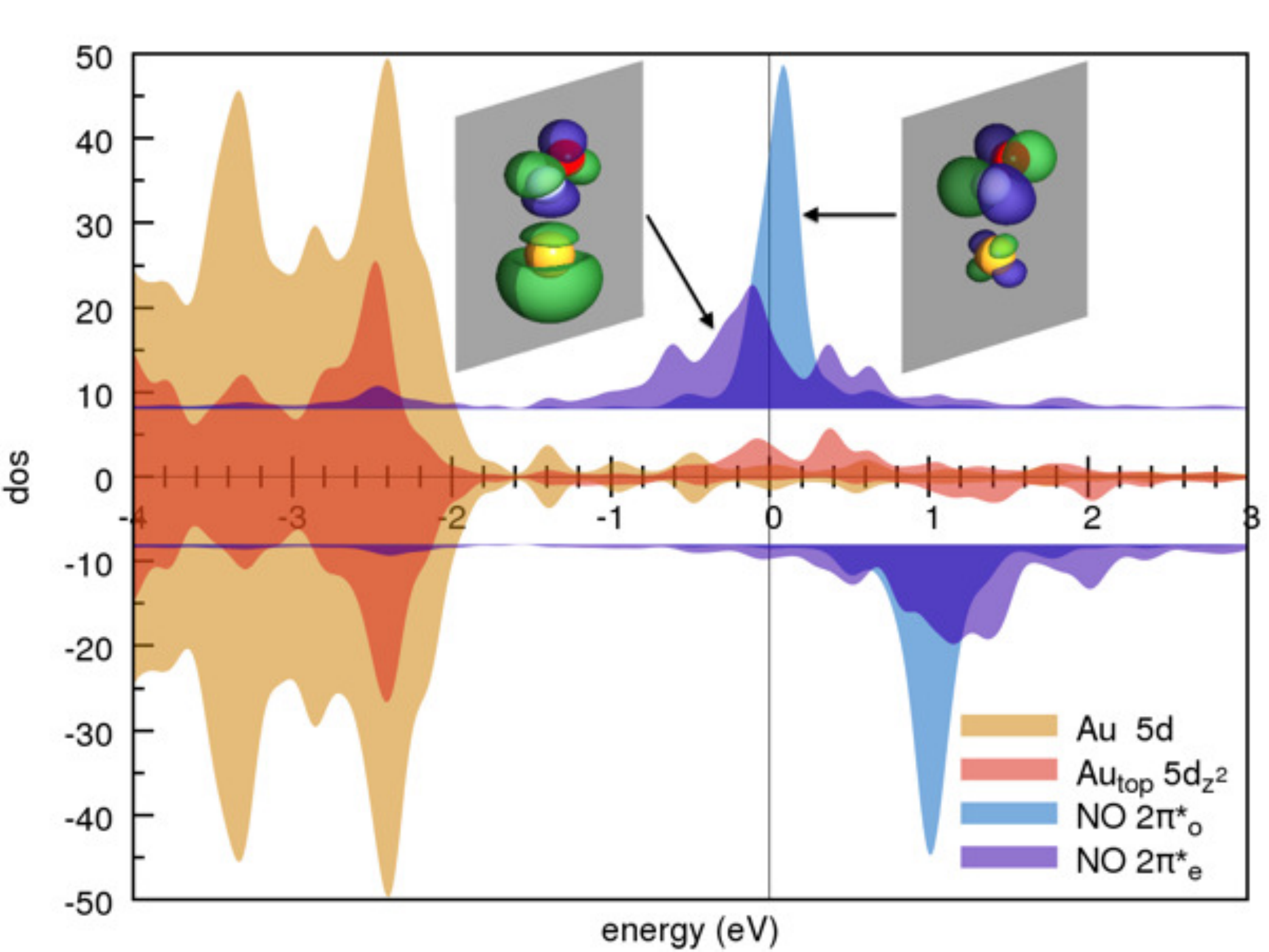}
\caption{\label{fig:dos-top} Projected density of states for NO/Au(111).  Spin up states are 
plotted as positive values, spin down as negative.  The density of states of the NO 2$\pi^*$ 
molecular orbitals (insets) are estimated from weighted sums of the projections on N and O 
$p$ orbitals.  The Fermi energy is indicated by a vertical line.   The 2$\pi^*$ states of NO 
and the $d_{z^2}$ states of the top-site Au atom, Au$_{\rm top}$, are shifted and scaled 
for visibility.}
\end{figure}
This is an important point -- the DFT calculation partitions the 
single unpaired electron approximately equally between the two $2\pi^*$ orbitals, despite 
a significant splitting of the orbital energy levels due to the large tilt angle.  Although the 
evidence of NO tilting in our STM images is not firmly established, its occurrence is 
convincingly stabilized by the resulting hybridization with the Au surface.  Of the total 
impurity magnetization $\sim$55\% originates from the more hybridized even $2\pi^*$ 
orbital with lobes in the Au$_{\rm top}$-N-O plane (see inset of Fig.~\ref{fig:dos-top}), 
and the remaining 45\% from the less hybridized odd $2\pi^*$ orbital normal to that 
plane.

To calculate the hybridization of the molecular orbitals with the gold surface, we use the 
information contained in the calculated phase shifts of gold conduction electrons scattering 
off the adsorbed molecule \cite{lucignano2009,baruselli2012a,baruselli2012b}.  The 
quantum mechanical scattering problem is solved numerically for the geometry in 
Fig.~\ref{fig:scat} using \textsf{pwcond} \cite{smogunov2004} (\textit{Materials and Methods}), which provides spin resolved phase shifts for the relevant symmetry channels.  
Here we focus on two symmetries, namely even ($e$) and odd ($o$) under reflection across 
the Au$_{\rm top}$-N-O plane, corresponding to the strongly and weakly hybridized NO 
$2\pi^*$ orbitals, respectively.  It should be noted for later discussion that the scattering 
problem involved only the Bloch states of bulk Au and not those of the surface states.

The next step of our theoretical protocol is to define an Anderson model Hamiltonian 
including channels of both symmetries
\begin{equation}
H = \epsilon_e n_e + \epsilon_o n_o + \sum_{\alpha=e,o} \sum_k  (V_{k\alpha} 
c_k^{\dag} c_{\alpha} + V_{k\alpha}^* c_{\alpha}^{\dag} c_k) + H_{\rm int},
\label{eq:H}
\end{equation}
where the subscript $\alpha=e$/$o$ stands for the $2\pi^*_{e/o}$ state, $V_{k\alpha}$ 
describes hybridization with gold conduction states, and $H_{\rm int}$ contains all of the 
interactions in the $2\pi^*$ manifold.  When the molecule is brought down to the surface 
in an upright configuration at the on-top, fcc or hexagonal close-packed (hcp) sites, the 
crystal field lowers the cylindrical symmetry of the isolated molecule to $C_{3v}$. Further 
tilting of the molecule at the on-top site breaks $C_{3v}$ symmetry and lifts the $2\pi^*$ 
degeneracy.  The splitting of the $2\pi^*_e$ and $2\pi^*_o$ levels and their unequal 
hybridization with gold conduction states, at first sight an irrelevant detail, is actually 
crucial to the Kondo physics, as we shall discuss below.  

The molecular orbitals hybridize with both surface and bulk states.  Surface state 
hybridization affects the electronic structure of adsorption but only weakly.  This is borne out 
in our slab calculations, where the adsorption energy converges already for 5 layers, while 
the energies of the surface states of the top and bottom of the slab---strongly split by their 
mutual interaction---do not converge until $\sim$23 layers \cite{forster2007}.  If surface 
state hybridization had a significant effect on the electronic structure, one would expect the 
adsorption energy to converge more slowly than it does.  Surface states decay exponentially 
inside the bulk, and therefore their contribution to the hybridization and phase shifts, and 
hence to the conductance anomalies, are invisible in our scattering calculations, which 
involve only bulk scattering channels.  We checked that the \textit{ab initio} estimates of 
the molecular orbital hybridization linewidths 
$2\Gamma_{\alpha} = 2\pi \sum_k |V_{k\alpha}|^2 \delta(\epsilon_k-\epsilon_F)$ are 
well converged with respect to the number of layers of the slab. 

The interaction Hamiltonian is now written in terms of the $2\pi^*$ states as
\begin{equation}
H_{\rm int} = U_e n_{e\uparrow} n_{e\downarrow} + U_o n_{o\uparrow} n_{o
\downarrow} + U_{eo} n_e n_o + J_H \textbf{S}_e \cdot \textbf{S}_o + W, 
\label{eq:Hint}
\end{equation}
where the $U$ terms are intra-channel and inter-channel Hubbard interactions, the $J_H$ 
term is their Coulomb ferromagnetic (Hund's rule) exchange interaction, and 
$W= W_{eo} c_{e\downarrow}^{\dag} c_{e\uparrow}^{\dag} c_{o\uparrow} c_{o
\downarrow} + W_{eo}^* c_{o\downarrow}^{\dag} c_{o\uparrow}^{\dag} c_{e
\uparrow} c_{e\downarrow}$ 
is a small but nonzero pair hopping term.  To fix the model parameters from the first 
principles DFT input, we require the scattering phase shifts of the model Hamiltonian at 
the Hartree Fock (mean-field) level to reproduce the phase shifts of the \textit{ab initio} 
scattering calculation for the full system.  The physical picture behind this type of matching 
is the local moment regime of the Anderson impurity model; since spin-polarized DFT 
and Hartree Fock provide comparable mean-field descriptions of the local moment regime, 
we can reliably infer the model parameters by requiring them to give the same phase shifts.
With four total phase shifts from the two channels and two spin polarizations, these 
conditions determine the four model parameters $\epsilon_e$, $\epsilon_o$, $U_e$ and 
$U_o$.  To fix the remaining parameters $U_{eo}$, $J_H$ and $W_{eo}$, we make use of 
relationships between the interaction parameters that are exact for the isolated molecule 
(\textit{SI Text}) and which should still apply to the adsorbed molecule owing to its gentle 
physisorption.  The resulting parameters (\textit{Table~S2}) provide all the ingredients we 
need for a many body calculation of the spectral function and Kondo conductance anomaly.

\bigskip
\noindent \textbf{NRG Results}

\begin{figure*}[t!]
\centering
\includegraphics[width=1.9\columnwidth]{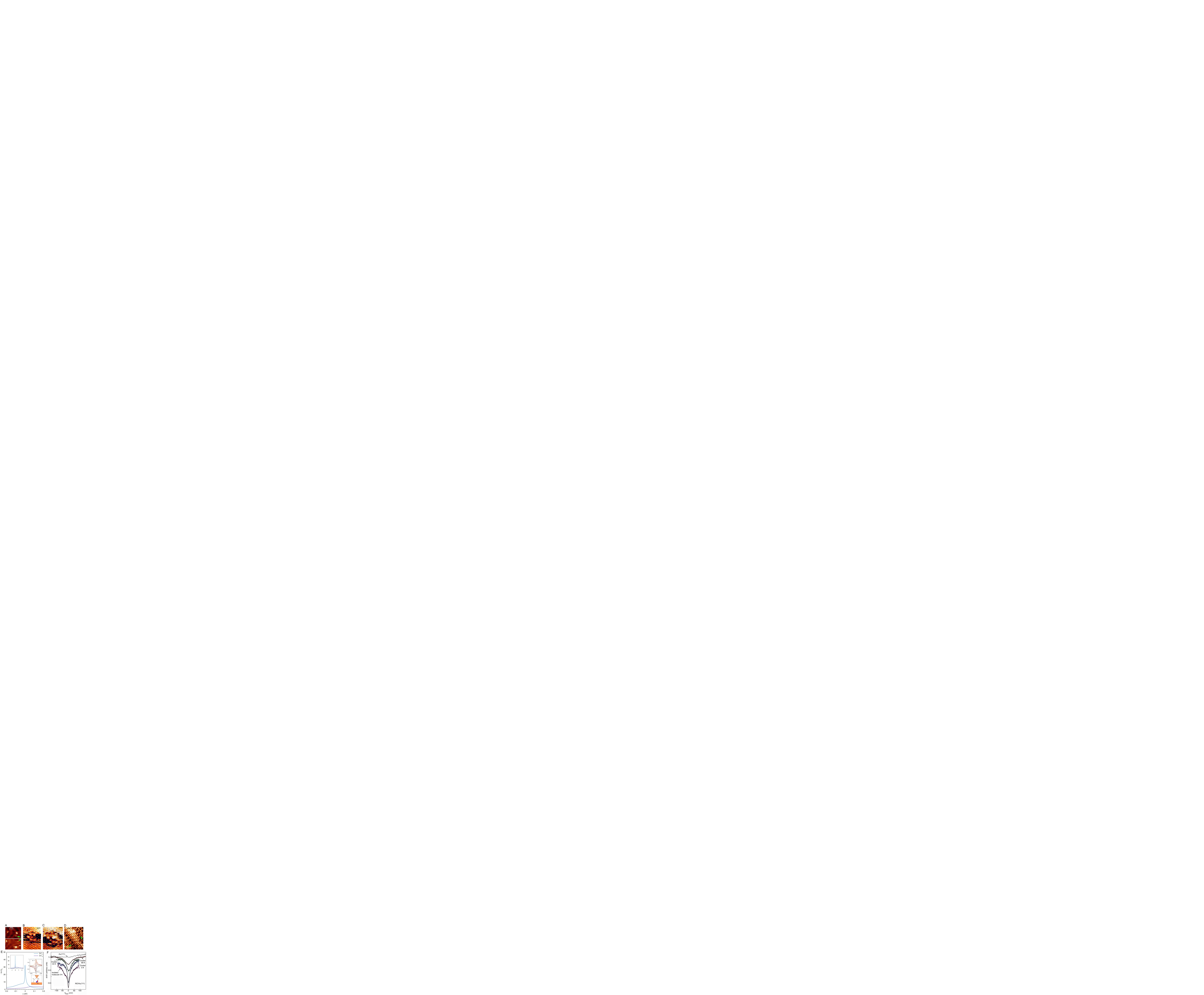}
\caption{\label{fig:spectra} STM/STS images and spectra, all obtained with clean Au tips 
except for B,C.  
(\textsl{A}) Isolated NO molecules (see arrows) and small NO clusters on Au(111) at 5 K. 
(\textsl{B,C}) Two small NO clusters formed from a few molecules (in the center of the 
figure) and substrate Au atoms (at the edges of the figure) imaged with a molecule on the 
STM tip; the Au lattice is marked by blue dots.  Most of the molecules are on-top.  The bar 
on the top is 0.5 nm.
(\textsl{D}) Filled state STM image ($V_{\rm bias}=-60$~mV) of an ordered single NO 
layer obtained after annealing at 70 K.  The layer is made up of rows of units with a 
dumbbell shape that is consistent with the shape of the filled $2\pi^*_o$ orbital.  The 
arrows point to two of these units.
(\textsl{E}) NRG spectral functions for the 2$\pi^*$ orbitals of NO at the on-top adsorption 
site calculated at 5~K with ab initio parameters from Table S2.  Insets: spectral function (left), 
Fano lineshapes (top right) and interfering tunneling paths $t_1$ and $t_2$ (bottom right). 
(\textsl{F}) Tunneling spectra of isolated molecules and small clusters at 5 K and clusters 
at 40 K and 68 K showing the Kondo dip. The spectra are shifted vertically for clarity.  The 
solid lines are fits with the Frota function \cite{frota1992}.}
\end{figure*}

\noindent The two-orbital Anderson model, Eq~(\ref{eq:H}), is solved by NRG 
\cite{bulla2008,zitko2006,zitko2009}; details are given in \textit{Materials and Methods}.
For tilted NO adsorbed at the on-top site, the solution indicates a competition between even 
and odd $2\pi^*$ orbitals.  The $2\pi^*_e$ is much more hybridized with the Au surface, 
whereas the $2\pi^*_o$ has a lower bare energy, and it is thus unclear at the outset
which of the two will host the Kondo resonance.  We find, somewhat surprisingly, that at the 
fixed point the least hybridized, odd orbital is the winner, its spectral function displaying the 
Kondo resonance shown in Fig.~\ref{fig:spectra}E.   By occupying the $2\pi^*_o$ more 
substantially than the $2\pi^*_e$, NRG inverts the mean-field orbital occupations initially 
obtained in DFT.  At $T=0$~K, the NRG orbital occupations are $n_o=0.83$ and 
$n_e=0.34$, while the DFT occupations were $n_o=0.44$ and $n_e=0.65$.  This 
exemplifies a situation of more general interest, showing that quantum fluctuations may 
cause one channel, even one predominant in mean field, to give way to another.

As a general note of caution we should stress that our Anderson model is in a regime where 
the parameter dependence of the Kondo temperature, lineshape, and orbital population 
balance, is rather critical.  Kondo temperatures are particularly difficult to predict, given 
their exponential dependence on parameters.  At the same time, the accuracy of the DFT 
electronic structure and phase shifts near $\epsilon_F$, which determine those parameters,
is limited by inaccuracies and approximations, including in particular self-interaction errors. 
The Kondo temperature increases by a factor of three if the hybridization is increased by 
20\% or the orbital energy splitting $\epsilon_e-\epsilon_o$ is decreased by only 50 meV.  
Further decreasing $\epsilon_e-\epsilon_o$ leads to a rather abrupt transition---via charge 
transfer from $2\pi^*_o$ to $2\pi^*_e$---to a state with a broad resonant level at 
$\epsilon_F$ rather than a Kondo resonance. We stick here to the results predicted by our 
straightforward protocol, without adjustments or data fitting.
 
\bigskip
\noindent \textbf{Experimental STM/STS Results}

\noindent  To verify the theoretical predictions just presented, we carried out STM 
measurements of NO adsorbed on Au(111).  Submonolayer NO was dosed on a 
reconstructed Au(111) surface at 30 K, and the sample was then cooled to 5 K.  STM 
images acquired with tunneling currents below 10$^{-11}$~A show that the molecules 
adsorb preferentially at the fcc elbows of the gold herringbone surface reconstruction as 
isolated units or small disordered clusters at low coverage---below about 0.05 
monolayers---and form large disordered clusters at higher coverage.  The images of small 
clusters in Figs.~\ref{fig:spectra}B-C indicate that the largest proportion of molecules 
adsorb on-top, in agreement with the theoretical prediction. The average nearest
neighbor NO-NO distance in the clusters is about 0.5 nm.  After annealing at 70 K, the 
molecules form ordered islands with a variety of metastable lattices with unit vectors 
between 0.35 and 0.60 nm. The angle between the unit vectors of the molecular lattices 
and those of the gold surface is between 0 and 30 degrees.  The structures are either 
incommensurate with the substrate or have a large unit cell with several inequivalent 
molecules, suggesting that the NO-NO interaction prevails over the NO-Au interaction. 
The NO lattice shown in Fig.~3D is formed by units that have a dumbbell appearance in 
the filled-state STM images.  This shape is consistent with the charge distribution of the 
$2\pi^*_o$ orbital of the tilted molecule (Fig.~\ref{fig:dos-top}) and supports the results 
of the NRG calculations that predict this state is preferentially filled.

STM spectra measured on isolated NO molecules, small disordered clusters, large 
disordered clusters and ordered NO lattices all display similar dip-shaped conductance 
anomalies centered at the Fermi level, as shown in Figure~\ref{fig:spectra}F.  For isolated 
molecules, whose mobility at 5 K was too high to acquire images with enough resolution to 
detect the predicted on-top adsorption state and tilt angle, the HWHM of the dip is about 12 
$\pm$ 4 meV at 5 K with an amplitude that is $\sim$25\% of the background conductance.  
For NO clusters and lattices, the dip HWHM is 16 $\pm$ 4 meV at 5 K and depends on the 
position of the molecule.  In addition to the zero bias dip, shoulders at $\pm 30$ meV are 
visible in the spectra taken over the isolated molecule but not over clusters 
(Fig.~\ref{fig:spectra}F).  We have not been able to determine whether these inelastic 
features are of vibrational or electronic origin, but the symmetric displacement of the 
shoulders with respect to zero bias, as well as their disappearance in clusters, is suggestive 
of a vibrational origin.  By DFT calculations we find vibrational eigenmodes with frequencies 
of 25 and 42 meV in the correct range. Spin transitions and anisotropy are not expected for 
a spin 1/2.  Hypothetically, electronic states might offer an alternative explanation, the 
lower Hubbard band of the $2\pi^*_o$ state, visible just below Fermi in the NRG spectra 
(Fig.~\ref{fig:spectra}E), corresponding to the left shoulder, and the empty $2\pi^*_e$ 
state forming a resonance above Fermi, to the right shoulder.  Altogether, an electronic 
explanation seems less likely.

No other sharp structures were observed between -0.8 and 0.8 eV from the Fermi level as 
shown in \textit{Fig.~S2} and described in \textit{SI Text}.  In particular, the large step at 
-0.45 eV corresponding to the bottom of the clean Au(111) surface state band is absent in 
the spectra measured over molecules, indicating that free surface state electrons are 
pushed out by the NO molecule, to become detectable again in spectra 1-2 nm away.  
Although surface states are repelled, nevertheless their tails are expected to extend up to 
and hybridize with the NO radical.

\begin{figure}[t]
\centering
\includegraphics[width=\columnwidth]{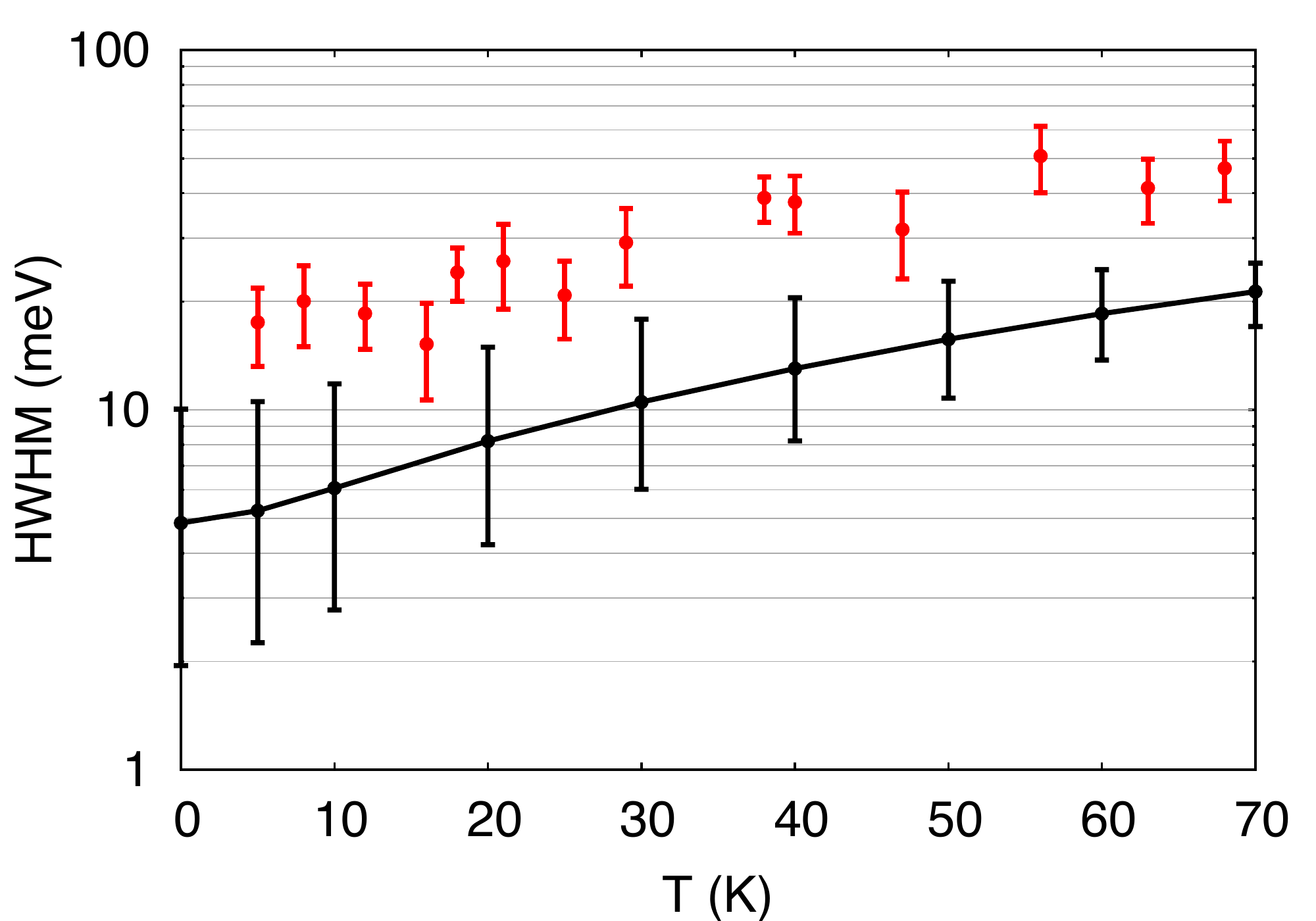}
\caption{\label{fig:temp-dep} Temperature dependence of the HWHM of the Kondo 
resonance from experiment (red) and NRG calculations (black) based on the ab initio 
parameters for the on-top adsorption geometry (\textit{Table~S2}).  Experimental error 
bars indicate the standard deviation calculated from a collection of spectra on different NO 
molecules.  Spectra were measured on at least 10 different molecules at each temperature.  
The error bars on the theoretical curve represent the uncertainty corresponding to a 
$\pm$~20 percent uncertainty in the hybridization linewidth; uncertainty from 
other parameters is not depicted.  Model parameter SDs are reported in Table 
S2.}
\end{figure}

To further confirm the Kondo nature of the observed zero bias STM anomaly, the 
temperature dependence of the half width at half maximum of the dip was measured and is 
shown in Figure~\ref{fig:temp-dep}.  For each temperature, the lineshape was fit with an 
asymmetric version \cite{prueser2012} of the Frota function \cite{frota1992} with
asymmetry parameter $\phi=\pi$.  By 70 K, the HWHM has grown to 40-50 meV and the 
amplitude has decreased from $\sim$25\% to $\sim$10\% of the background.  Due to the 
high mobility of the isolate molecules at high temperature and the ensuing difficulty of 
acquiring spectra for them, all data presented for $T>5$ K were taken on NO islands.  The 
temperature induced broadening and attenuation of the dip is intrinsic because it is about a 
factor of five larger than the effect of the change of the experimental resolution of STS at 
high temperature related to the broadening of the Fermi distribution of the tip, 
corresponding to a HWHM of about 10 meV at 68 K.  The large intrinsic broadening rules 
out inelastic tunneling (vibrational or electronic excitations) as the main origin of the dip.  
The Kondo temperature $\sim$43 K, extracted from the low temperature limit of the Frota 
linewidth parameter $\Gamma_K$, is about a factor of three larger than our unbiased 
theoretical prediction. This level of discrepancy is not surprising in view of the exponential 
sensitivity of the Kondo temperature to the impurity parameters, and of the self-interaction 
errors generally affecting the first principles DFT calculations.

Actually, a theoretical underestimate of the Kondo temperature is not unexpected, since the 
hybridization with gold surface states, as was mentioned above, is not completely 
accounted for in our calculations.  Despite these deficiencies, the overall, unbiased 
theoretical results are gratifyingly reproduced, and so is the evolution of the zero bias 
anomaly with temperature, as shown in Fig.~\ref{fig:temp-dep}. 
The experimental offset 
of the center of the dip from the Fermi energy, less than 2 meV, is much less than the 
width, indicating that the parent level of the Kondo resonance is indeed nondegenerate. 
That provides an indirect and yet strong confirmation that the NO molecule is tilted, for if 
the $2\pi^*$ levels were degenerate the Kondo peak would be shifted above the Fermi 
energy to satisfy the Friedel sum rule \cite{choi2005}. The data do not clearly establish if 
the symmetry of the Kondo orbital is odd as predicted, and that remains to be verified, 
e.g., by photoemission spectroscopy.

Finally, the observed Fano conductance lineshape (inset Fig. 3E), a $q\simeq 0$ dip rather
than $q\simeq\infty$ as predicted by the impurity spectral function, indicates that direct 
tunneling into the molecular $2\pi^*$ states ($t_2$ in inset Fig. 3E) is dominated by 
another tunneling channel ($t_1$).  As STS indicates, the NO molecule is fully immersed in 
the Au(111) surface states, whose wave functions extend far out into the vacuum, reaching 
out to the STM tip.  Because surface states do not represent a bulk screening channel, they 
are not included in our scattering calculations. Since the surface states are nonetheless 
involved and provide a free electron reservoir for Kondo screening, their local density of 
states near the molecule will show a dip at the Fermi energy, complementary to the 
calculated bare NO spectral function peak.  As observed in the Kondo STM conductance 
anomalies of many other adsorbed magnetic atoms, and as generally discussed by 
Ujsaghy et al. \cite{ujsaghy2000}, $dI/dV$ spectra are strongly modified, in our case very 
plausibly dominated, by direct tunneling into nearby metal surface states, specularly 
reversing the intrinsic Kondo anomaly. 

\bigskip
\noindent \textbf{Conclusions and Outlook}

\noindent We have described how the Kondo parameters and resulting zero bias STM 
conductance anomaly of an adsorbed molecular radical can be predicted from first 
principles, with features that compare reasonably well with a subsequent experiment. 
To do so, we employed NO as the simplest spin 1/2 molecular radical capable of forming 
a Kondo screening cloud when adsorbed on a gold surface.  The discrepancies between 
the calculated Kondo temperature and lineshape, which we did not try to amend in any 
way, and the measured ones are quite instructive, highlighting in particular the need to 
fully incorporate the metal surface states in future calculations. The delicate 
renormalization group flow towards the least hybridized, odd symmetry $2\pi^*_o$ 
state also appears as an instructive many body effect, and remains a prediction to be 
verified.  The magnetic field splitting of the Kondo anomaly, presently academic in view 
of the large value of $T_K$, could easily be calculated if need be. 

The protocol implemented here for NO/Au(111) in a specific STM geometry is of more 
general applicability and can be applied to different molecular radicals and different STM 
and break junction geometries, where the influence of structural or mechanical 
deformations could be explored.  As an overall result, the DFT+NRG approach demonstrates 
enough predictive power to be useful in the \textit{a priori} evaluation of the conductance 
characteristics of molecular radical nanocontacts, thus providing a theoretical asset of 
considerable technological relevance.

\bigskip
\noindent \textbf{Materials and Methods}

\noindent Density functional theory calculations in the generalized gradient approximation of Perdew, Burke and Ernzerhof \cite{perdew1996} were performed with Quantum Espresso 
\cite{giannozzi2009}, a plane wave pseudopotential electronic structure package.  The 
NO/Au(111) adsorption geometry was determined using the slab method in a 3$\times$3 
surface supercell with a vacuum layer of 16.9~\AA.  Scattering calculations were carried out 
with pwcond \cite{smogunov2004} in a 3$\times$3 supercell with a vacuum layer of 
8.4~\AA, in correspondence with the experimental tip height of $\sim$10~\AA.  Brillouin 
zone integrations were performed on a 6$\times$6$\times$2 $k$-point mesh with a 
smearing width of 0.002-0.010 Ry.  Plane wave cut-offs were 30 Ry for the wave function 
and 360 Ry for the charge density.  As detailed in (SI Text), Hubbard interactions were 
applied to the $d$ states of Au (U=1.5 eV) and the $p$ states of N and O (U=1.0 eV).

Numerical renormalization group calculations were carried out with NRG Ljubljana 
\cite{zitko2006} employing the $z$-averaging technique \cite{zitko2009}, the full density 
matrix approach \cite{weichselbaum2007} and the self-energy trick \cite{bulla1998}.  
The logarithmic discretization parameter was chosen to be $\Lambda=4$, and a maximum 
of 2000 states (not counting multiplicities), corresponding to roughly 5000 total states, 
were kept at each iteration.  Spectral functions were obtained by log-gaussian broadening 
\cite{bulla2001} and at finite temperature with a kernel that interpolates between a log-
gaussian at high energy and regular gaussian at low energy \cite{weichselbaum2007}.

The dI/dV spectra were measured with a clean Au tip by the lock-in technique applying a 
3 meV modulation to the bias voltage and using a maximum current of 
$3\times 10^{-11}$~A to avoid tip induced modification of the adsorption geometry.  NO 
molecules were easily displaced at higher tunneling currents at 5 K and were mobile at 
temperatures above 20-30 K.  The gold lattice and molecules could be resolved 
simultaneously only once, Figs.~3B and C, when an unknown molecule adsorbed on the 
tip.  The horizontal stripes in Figs.~3B and C are caused by temporary detachments or shifts 
of this molecule.  The broadening of the tunneling spectra caused by the finite modulation 
voltage is taken into account in the fit of the measured spectra by calculating the convolution 
of the Frota function with $\sqrt{1-(V/V_{\rm pp})^2}$, where $V_{\rm pp}$ is the peak-
to-peak value of the modulation voltage and $V$ is the bias.  This function represents the 
response of the STS spectroscopy with a sinusoidal modulation of the bias and a lock-in 
amplifier to a delta function-like density of states.  The effect of the thermal broadening of 
the Fermi distribution in the tip at high temperature was taken into account by computing 
the convolution of the Frota function with the derivative of the Fermi distribution.

\begin{acknowledgments}
We are grateful for the use of the NRG Ljubljana code and the high performance computing 
resources of CINECA.  Work was partly sponsored by contracts PRIN/
COFIN 2010LLKJBX 004 and 2010LLKJBX 007, Sinergia CRSII2136287/1 and advances of 
ERC Advanced Grant 320796 -- MODPHYSFRICT.
\end{acknowledgments}

%%%%%%%%%%%%%%%%%%%%%%%%%%%%%%%%%%%%%%%%

\newpage
\clearpage

\renewcommand{\thefigure}{S\arabic{figure}}
\renewcommand{\thetable}{S\arabic{table}}
\renewcommand{\theequation}{S\arabic{equation}}

\setcounter{figure}{0}
\setcounter{table}{0}
\setcounter{equation}{0}

\begin{flushleft} 
{\Large \bf \textsf{Supporting Information}}\\
\bigskip 
{\normalsize \bf \textsf{R. Requist, et al.}}
\bigskip
\end{flushleft}

\noindent {\bf \textsf{SI Text}}
\medskip

\noindent {\bf \textsf{First principles electronic structure calculations}}\\
Density functional theory calculations of NO adsorption were performed using the slab 
method with 3-7 gold layers and a $3\times 3$ hexagonal supercell of the unreconstructed 
Au(111) surface, corresponding to 1/9 monolayer coverage.  This cell was large enough to 
reduce the interaction between periodic images of the NO molecule to a negligible level.  
Selected calculations were repeated with $2\times 2$, $2\sqrt{3}\times\sqrt{3}$ and 
$2\sqrt{3}\times2\sqrt{3}$ cells to verify convergence.  The unreconstructed Au(111) 
surface is expected to be a good approximation to the face-centered cubic (fcc) regions of 
the well-known $22\times \sqrt{3}$ ``herringbone'' reconstruction, where the 
experimental measurements were performed.  For calculations of the adsorption energy, 
the vacuum layer was set to 16.9~\AA, while for calculations of the scattering phase shifts, 
it was reduced to 8.4~\AA, corresponding to the experimental tip height.  The latter value 
was large enough that the shape of the tip (cf. Fig.~1 of the main article) had a negligible 
effect on the adsorption geometry and hybridization linewidths.  Hence, the scattering 
calculations were performed without any model for the tip, essentially using the bottom of 
the slab as a broad distant probe. 

The unreconstructed Au(111) surface presents 4 distinct high-symmetry adsorption sites -- 
fcc, hexagonal close packed (hcp), bridge and on-top.  For each site, all the coordinates of 
NO and the two highest layers of gold were fully optimized.  By subsequently applying 
constraints to the Au atoms, it was found that only the relaxation of the Au atoms nearest 
to N had a significant effect on the results, and therefore in the scattering calculations we 
fixed all Au atoms to their bulk positions, except the 1, 2, or 3 Au atoms nearest N in the 
case of top, bridge, and hollow site adsorption, respectively.  At the top site, the Au atom 
directly beneath N, which we label as Au$_{\rm top}$, is pulled out of the surface by 
0.11~\AA, enhancing the interaction between the NO $2\pi^*$ molecular orbitals and an 
$spd$ hybrid orbital (mainly $d_{z^2}$) of Au$_{\rm top}$. 

Table S1 summarizes our \textit{ab initio} results for gas phase NO and 
NO/Au(111), including the adsorption energy $E_{\rm ads}$, magnetic moment $\mu$, 
NO bond length $d_{\rm NO}$, AuN bond length $d_{\rm AuN}$, the stretching frequency 
$\nu_{NO}$, and electric dipole moment calculated with GGA+$U$. The magnetic moment 
is reported for the entire cell, including the small induced magnetization on gold.  The 
surface-adsorbate electric dipole moment was calculated by applying a sawtooth-shaped potential along the $z$ direction.  At the top site, NO was found to tilt away from the upright configuration, giving a Au$_{\rm top}$-N-O angle of 122.4$^{\circ}$, in agreement with earlier calculations \cite{gajdos2006}.  The tilting, which accounts for a large part of the adsorption energy, is crucial for the present study because it alters the Kondo physics by lifting the degeneracy of the $2\pi^*$ orbitals.  The molecule tilts toward one of the six nearest neighbor surface gold atoms of Au$_{\rm top}$. For definiteness, we shall take the molecule to tilt in the positive $x$ direction, so that the tilt plane is the $xz$ plane.  The degenerate $2\pi^*$ orbitals split into two orbitals that we label as $2\pi^*_e$ and $2\pi^*_o$ according to their symmetry (even or odd) with respect to reflection through 
the tilt plane.  Full geometry optimization shows that the tilt plane undergoes a small azimuthal rotation of 8.6$^\circ$ around the surface normal.   The order of magnitude of 
the calculated adsorption energies agrees with an estimate of 400 meV from temperature programmed desorption spectroscopy \cite{mcclure2004} and the results of a 
comprehensive study \cite{gajdos2006} of NO adsorption on metal surfaces, although the 
site dependence differs.  We are not aware of any further experimental data for NO/Au(111), but our calculations of the molecule in vacuum (Table S1) are in good 
agreement with gas phase measurements.  The calculated ionization energy of NO, 
9.007~eV, is also fairly close to the experimental value 9.27483 $\pm$ 0.00005~eV \cite{reiser1988}.  The work function of gold, calculated with a 24-layer slab with 
24~\AA~of vacuum, is 5.19~eV and compares favorably with the experimental value 
5.31~eV \cite{michaelson1977}.

In view of the weak adsorption energy, selected calculations were performed with the 
following four functionals in order to judge the sensitivity of the results to the choice of 
exchange-correlation functional:  i) the local (spin) density approximation (LDA) of Perdew 
and Zunger~\cite{perdew1981}, ii) the generalized gradient approximation (GGA) of 
Perdew, Burke and Ernzerhof~\cite{perdew1996}, iii) the Heyd-Scuseria-Ernzerhof (HSE) 
hybrid functional \cite{heyd2003}, and iv) a modified version \cite{sabatini2013} of the 
Vydrov-Van Voorhis (VV10) functional \cite{vydrov2010} describing van der Waals 
interactions.  The LDA approximation, known to be overbonding for molecules on surfaces, 
was found to be inadequate, causing NO to demagnetize.  The VV10 functional increased 
the adsorption energy uniformly for all sites, yielding 540, 380 and 360~meV for the on-
top, bridge and fcc sites, respectively. Hybrid functionals have been found to give good 
results for the adsorption of small molecules on metal surfaces \cite{wang2007}; however, 
since they open a small gap at the Fermi energy in metals, they are potentially problematic 
when the molecular levels lie close to the Fermi energy, as they do in our case.  Moreover, 
such a gap would cause artifacts in the calculation of the scattering phase shifts.  For these 
reasons, all of the results reported in the main article were obtained with the GGA and 
GGA+$U$.  

Hubbard interactions $U_{\rm N}=U_{\rm O}=1$~eV were applied to the N and O $p$ 
orbitals in the GGA+$U$ scheme in order to stabilize the magnetic moment.  Similar values 
have been used for CO to correct the adsorption site preference on Pt(111) 
\cite{kresse2003} and Cu(111) and Cu(001) \cite{gajdos2005}.  We also used Hubbard 
interactions to correct the energy of the Au $d$ bands, which in LDA and GGA 
\cite{takeuchi1991,mazzarello2008} are too high compared with angle-resolved 
photoemission spectroscopy \cite{kevan1987}.  We found that the value 
$U_{\rm Au}=1.5$~eV shifts the fully occupied $d$ bands down rigidly by $0.5$~eV, bringing them into agreement with experiment, while leaving the $sp$ bands virtually unchanged.  This value of $U_{\rm Au}$ was adopted in calculating all quantities in 
Table S1.  The GGA+$U$ scheme has been used in a similar way to correct 
the $d$ states of Ni in a study of the adsorption of CO and NO/NiO(100) \cite{rohrbach2004,rohrbach2005}.  

The projected density of states for the isolated molecule, clean surface and the combined 
surface-adsorbate system are shown in Figure~\ref{fig:dos-top}.  The most important observation is that the degeneracy of the $2\pi^*_e$ and $2\pi^*_o$ orbitals is broken by the tilting of the molecule, yet both remain partially occupied.  The tilting of the molecule away from the upright configuration increases the hybridization of the $2\pi^*_e$ orbital and decreases the hybridization of the $2\pi^*_o$ orbital.  As reported in 
Table S2, the hybridization $\Gamma_e$ becomes fives times as large 
as $\Gamma_o$ for the optimal tilt angle of nearly 60 degrees.  

Despite the substantial tilting of the molecule, there is only a relatively weak symmetry 
breaking in the $2\pi^*$ orbital occupations.  One might have expected the $2\pi^*_e$ 
orbital to be nearly empty, since its bare energy is pushed up by as much as $0.24$~eV 
due to its antibonding interaction with the $spd$ hybrid orbital of Au$_{\rm top}$.  
Instead, it is even more occupied than the $2\pi^*_o$ orbital.  The $2\pi^*_e$ and 
$2\pi^*_o$ molecular orbitals have fractional occupation numbers, $n_e= 0.65$ and 
$n_o= 0.44$, adding up to slightly more than 1, consistent with modest charge transfer 
from the surface to the molecule.  The degeneracy of the $1\pi$ orbitals is also lifted by 
the tilting, but they split in the opposite direction because they lie below rather than above the $d$ states in energy. The lack of distinct symmetry breaking in the $2\pi^*$ orbital occupations might be a spurious result, possibly caused by self-interation error, or it might 
be a genuine consequence of orbital fluctuations.  Hubbard interactions in nitrogen and oxygen are unable to induce a stronger symmetry breaking even though they increase the 
$2\pi^*_e$/$2\pi^*_o$ splitting for the spin and orbital symmetry-broken solution in 
vacuum.  It might be possible to achieve orbital symmetry breaking by applying 
Hubbard-type interactions to molecular orbitals rather than atomic orbitals 
(cf.~Ref.~\cite{kresse2003}) or in an approach with inherently less self-interaction error 
such as reduced density matrix functional theory, but we do not pursue these approaches here.  

The Blyholder \cite{blyholder1964} and Hammer-Morikawa-Norskov \cite{hammer1996} 
models, which describe the adsorption of CO on metals in terms of $\sigma$ donation and 
$\pi$ back-bonding, provide a starting point for understanding the bonding interactions 
between NO and the gold surface, however $\sigma$ and $\pi$ are no longer proper 
symmetries due to the tilting of the molecular axis.  Modest back-bonding interactions are 
visible at the top of the $d$ band ($\sim$2.5 eV below the Fermi energy) in 
Fig.~\ref{fig:dos-top}, coinciding with peaks in the Au$_{\rm top}$ $d_{z^2}$ states.  
The strong participation of the Au$_{\rm top}$ $d_{z^2}$ states in the $2\pi^*_e$, 
$5\sigma$, $1\pi_e$ and $4\sigma$ molecular resonances is an important feature of the 
bonding interaction.  There are important differences between the adsorption of NO and CO 
on noble metals, since NO is an open-shell molecule.
\smallskip

\noindent {\bf \textsf{Constructing an Anderson impurity model}}\\
\noindent An Anderson model representing the hybridization of the $2\pi^*$ molecular 
orbitals of NO with the Au(111) surface was defined in Eq.~\textbf{1} of the main article.  
In specifying the interaction Hamiltonian $H_{\rm int}$, it is convenient to start from the 
molecule in vacuum.  Since the isolated molecule has cylindrical symmetry, the interactions 
in the $2\pi^*$ sector depend on only two independent parameters and can be expressed 
as 
\begin{equation}
H_{\rm int} = \frac{V}{2} N(N-1) - 2J \mathbf{S}^2 + \frac{J}{2} L_z^2, 
\label{eq:Hint}
\end{equation}
where $N$ is the total number of electrons, $\mathbf{S}$ is the total spin operator, and 
$L_z$ is the total $z$ component of angular momentum.  The two degenerate $2\pi^*$ 
states, formed from the $p_x$ and $p_y$ orbitals of N and O, can be chosen to be 
eigenstates of $L_z$, i.e.~$|m=\pm 1\rangle$.  The $|m=0\rangle$ state is not considered 
because the $p_z$ orbitals, being involved in $\sigma$ bonding, are far lower 
($\sim$7.5~eV) in energy.  Since we will be considering the symmetry breaking caused by 
tilting, we express Eq.~\eqref{eq:Hint} in terms of the $2\pi^*_x$ and $2\pi^*_y$ states, 
which have nodes in the $yz$ and $xz$ planes, respectively.  Using the relations 
$c_{m=1} = (-c_x - i c_y)/\sqrt{2}$ and $c_{m=-1} = (c_x - i c_y)/\sqrt{2}$, we find
\begin{equation}
H_{\rm int} = U_x n_{x\uparrow} n_{x\downarrow} + U_y n_{y\uparrow} 
n_{y\downarrow} + U_{xy} n_{x} n_{y} + J_H \textbf{S}_x \cdot \textbf{S}_y + W + 
\beta N, \label{eq:Hint:alt}
\end{equation}
where the $U$ terms are on-site and inter-site Hubbard interactions, the $J_H$ term is a 
Hund interaction ($\mathbf{S}_{\alpha}$ is the spin operator for the $2\pi_{\alpha}^*$ 
state, \textit{not} the $\alpha$-component of spin) and $W$ is a double hopping term 
$W = W_{xy} c_{x\uparrow}^{\dag} c_{x\downarrow}^{\dag} c_{y\downarrow} 
c_{y\uparrow} + W_{xy} c_{y\uparrow}^{\dag} c_{y\downarrow}^{\dag} 
c_{x\downarrow} c_{x\uparrow}$.  The parameters are uniquely determined by $(V,J)$ 
according to the formulas $U_x = U_y = V + 3J$, $U_{xy} = V - J/2$, $J_H = -6J$, 
$W_{xy} = -J$ and $\beta = -J/2$.  

When the molecule is brought down to the surface in an upright configuration at the on-top, 
fcc or hcp sites, the crystal field lowers the cylindrical symmetry to $C_{3v}$ symmetry.  
The degeneracy of the $2\pi^*$ states is preserved and Eqs.~\eqref{eq:Hint} and 
\eqref{eq:Hint:alt} remain exact.  The tilting of the molecule breaks the symmetry and lifts 
the $2\pi^*$ degeneracy.  The $2\pi^*_x$ orbital evolves into the even $2\pi^*_e$ orbital 
(see Fig.~2 of the main article) that hybridizes strongly with the surface.  The $2\pi^*_y$ 
orbital becomes the odd $2\pi^*_o$ orbital with much weaker hybridization.  
Equation~\eqref{eq:Hint:alt} remains a valid description of interactions in the $2\pi^*$ 
sector; however, since the $2\pi^*_e$ and $2\pi^*_o$ orbitals are hybridized with the 
surface, the interaction parameters will no longer have exactly the same relationship with 
$(V,J)$ that they have in cylindrical symmetry.  

For tilted NO at the on-top site, there were not enough \textit{ab initio} data to fit all of the 
parameters in Eq.~\eqref{eq:Hint:alt}, so we have taken the following strategy.  We have 
fit the parameters that are most sensitive to the tilting, namely $\epsilon_{\alpha}$ and 
$U_{\alpha}$, by matching the spin-symmetry broken mean-field 
$\epsilon_{\alpha\sigma}$ of the Anderson model to $\epsilon_{\alpha\sigma}^{DFT}$ 
inferred from the resonances in the \textit{ab initio} phase shifts.  The hybridization 
linewidths $2\Gamma_{\alpha}$ were calculated by fitting the resonances in the scattering phase shifts to the following functional form
\begin{align}
\eta_{\alpha\sigma}(\epsilon) = \frac{\pi}{2} + \arctan \frac{\epsilon-\epsilon_{\alpha
\sigma}}{\Gamma_{\alpha}} + \delta_{\alpha}, 
\label{eq:phaseshift}
\end{align}
where $\delta_{\alpha}$ is a constant shift attributable to potential scattering.  An example of the fitting is shown in Fig.~\ref{fig:phaseshift}.  The Hund interaction $J_H$ was set equal to its value for gas phase NO, and the remaining parameters $U_{eo}$ and $W_{eo}$ were assumed to have the same relationship to $V$ and $J$ that they have in cylindrical symmetry (as in gas phase).  In this way, all model parameters can be fit reliably, and we obtain the results in Table S2.

For upright NO at the bridge site, the symmetry is lowered to $C_{2v}$.  The 
$2\pi^*_{e/o}$ state is defined to be the state that is even/odd with respect to the plane 
containing NO and the two nearest neighbor Au atoms.  The interaction parameters were 
approximated the same way as for the on-top site.

Symmetry between the $2\pi^*_{x/y}$ states is preserved for upright NO at the fcc site.  The interaction parameters were approximated the same way as for the on-top site; however, since the occupied (majority spin) level $\epsilon_{\alpha\uparrow}^{DFT}$ is degenerate and therefore pinned to the Fermi energy, we obtained a more stable fit for 
$\epsilon_{\alpha}$ by requiring charge consistency between DFT and NRG, 
i.e.~$n_{\alpha}^{DFT}=n_{\alpha}^{NRG}$.
\smallskip

\noindent {\bf \textsf{Scanning tunneling spectroscopy}}\\
\noindent The densities of states measured with a clean Au tip over NO molecules and over 
the clean Au(111) surface several nm away from the molecules is shown in Fig. \ref{fig:spectra:extended} for an extended energy range around the Fermi energy.  The 
step at about -0.45 eV on the gold surface marks the bottom of the surface state band \cite{chen1998}.  The surface states cannot be detected on top of the NO molecules and gradually vanish when the tip approaches the molecules. The only sharp feature in the NO spectrum above the noise level is the zero-bias dip.  The broad and weak bumps at about -0.15 eV and 0.3 eV approximately coincide with the shoulders of the spectral functions of Fig.~3F of the main article.

\begin{figure*}[h!]
\centering
\includegraphics[width=\textwidth]{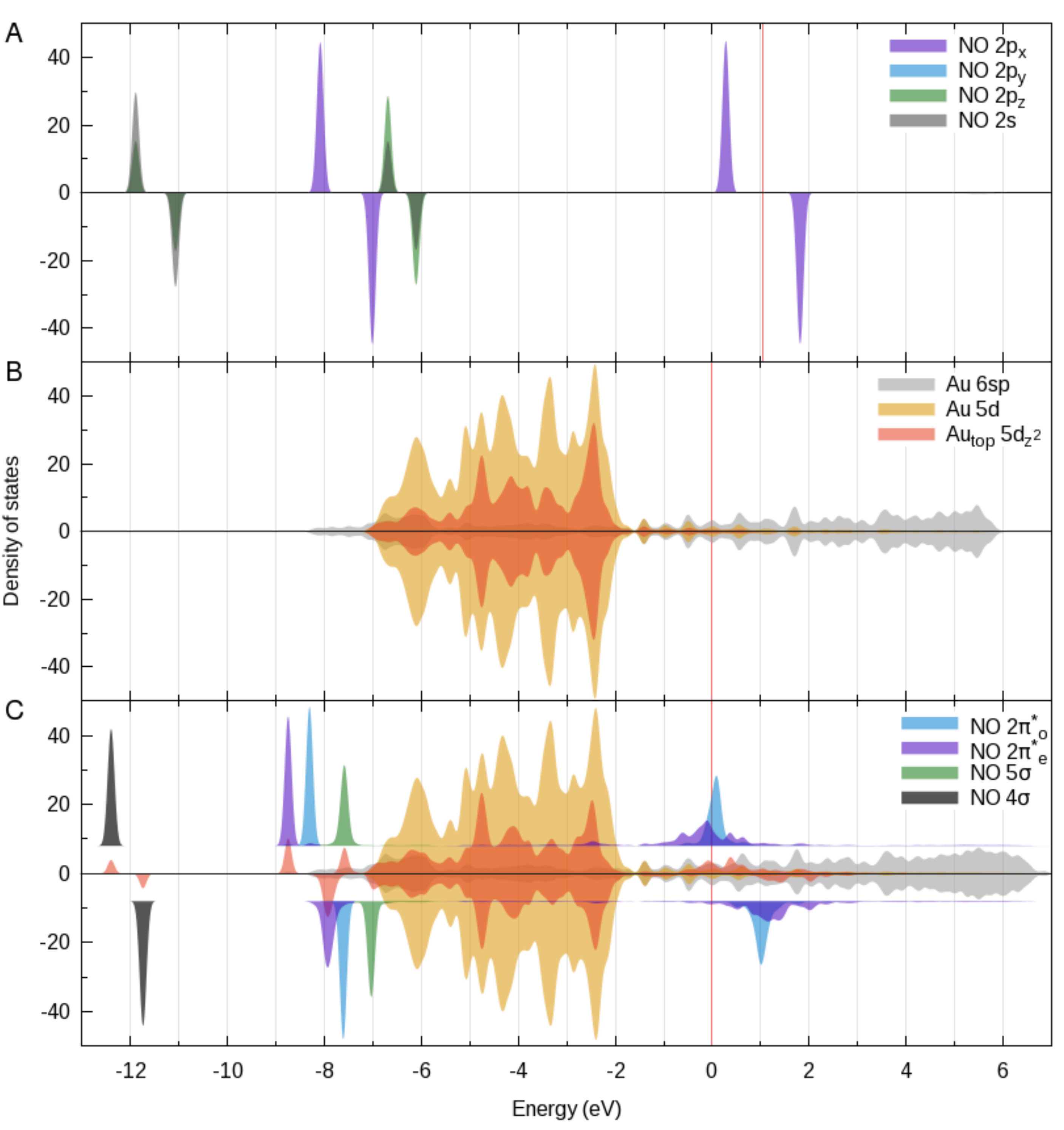}
\caption{Projected density of states of (\textsl{A}) Gas phase NO, (\textsl{B}) 3-layer Au(111) slab, and (\textsl{C}) NO/Au(111) slab system.  Energies have been shifted so as to align the vacuum levels of all systems; vertical red lines indicate the respective Fermi energies.  Artificial gaussian broadening of 0.10 eV is used.  NO $2p_x$ and $2p_y$ orbitals are degenerate in gas phase.  The density of states of the 2$\pi^*$ orbitals are calculated from weighted sums of N and O $p_x$ and $p_z$ orbitals, corresponding to the tilting.}
\label{fig:dos-top} 
\end{figure*}

\begin{figure}[h]
\centering
\includegraphics[width=\columnwidth]{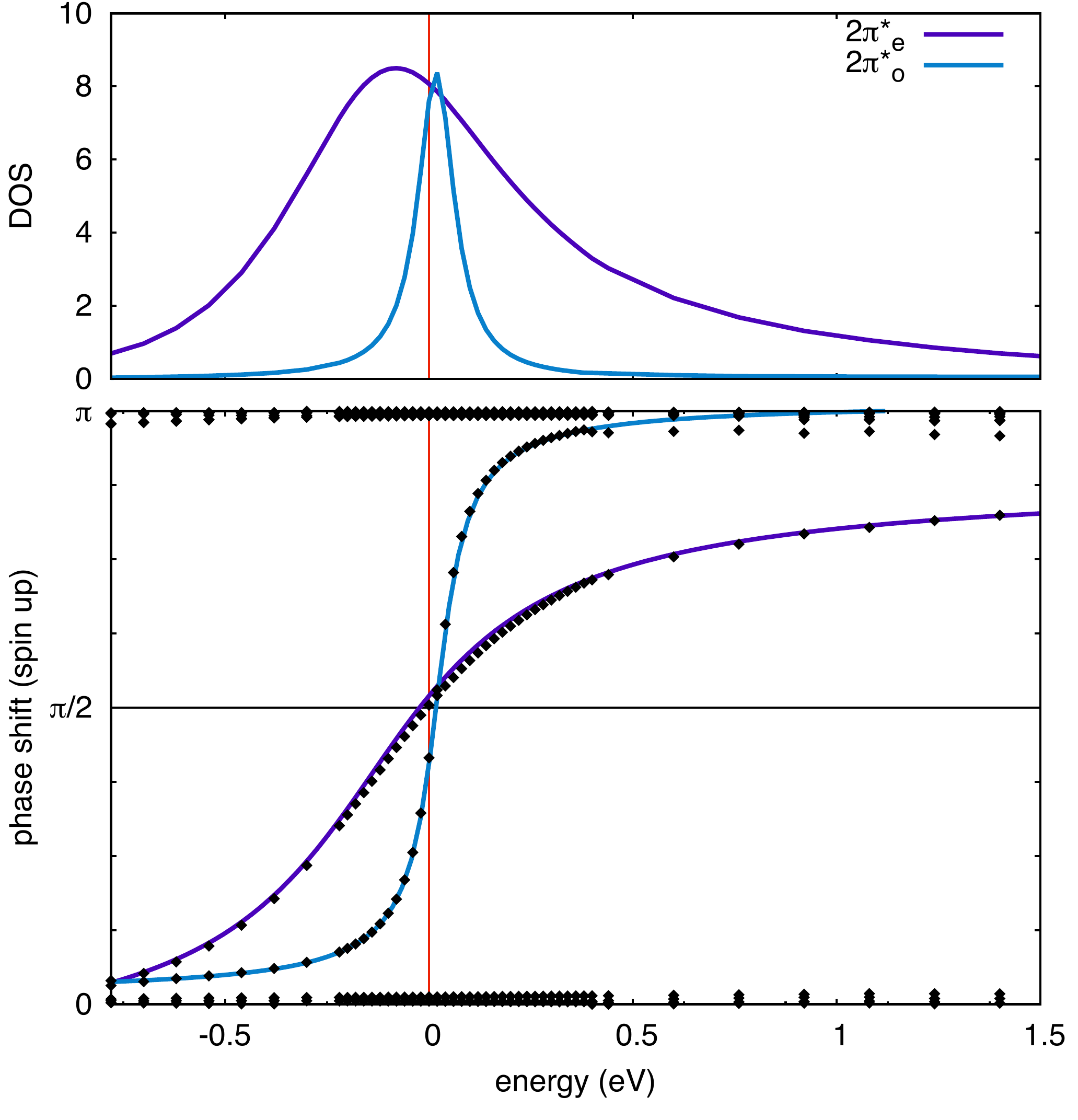}
\caption{Example of resonances in the scattering phase shifts and fits to Eq.~\ref{eq:phaseshift}.}
\label{fig:phaseshift}
\end{figure}  

\begin{figure}[h]
\centering
\includegraphics[width=\columnwidth]{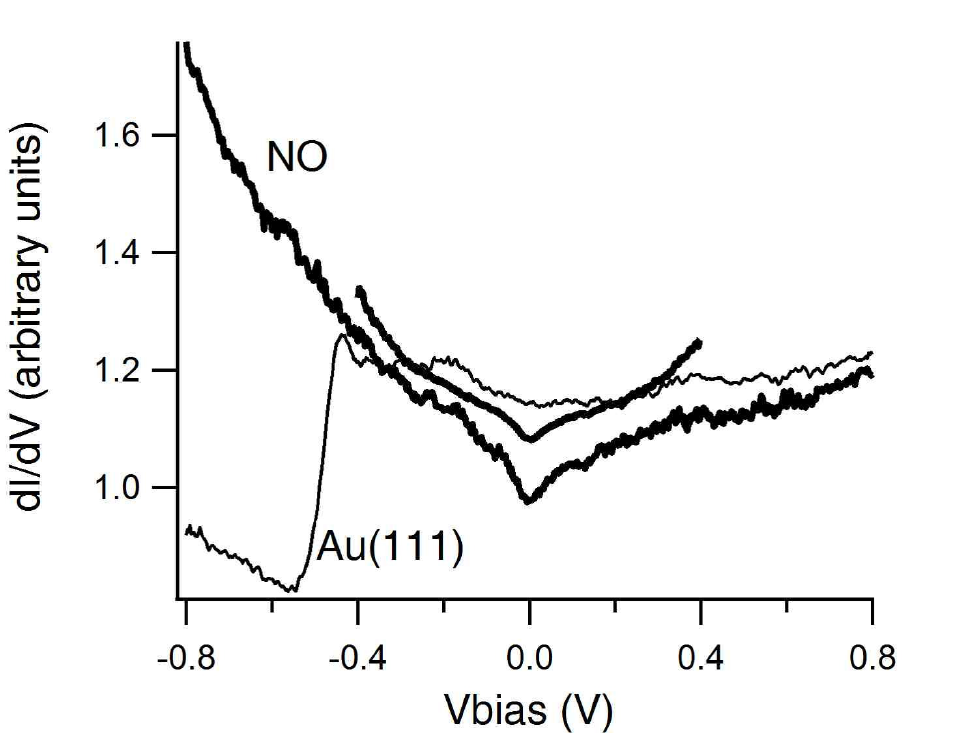}
\caption{STS spectra taken over NO molecules at 70 K (thick lines) and over the clean Au(111) surface a few nm away at 5 K (thin line).}
\label{fig:spectra:extended}
\end{figure}  

\begin{table*}[h!]
\centering
\caption{Ab initio properties of gas phase NO and NO/Au(111)}
\begin{tabular*}{0.8\textwidth}{@{\extracolsep{\fill}}llllllll} 
site   & config.  & $E_{\rm ads}$ (eV) & $\mu$ ($\mu_B$) & $d_{\rm NO}$ (\AA) & 
$d_{\rm AuN}$ (\AA) & $\nu_{\rm NO}$ (cm$^{-1}$) & elec.~dipole~(D) \\ \hline
gas phase    & ---     &  ---      & 1.00 & 1.1662 & ---    & 1885 & 0.109 \\  
&  &  &  & (1.148)\tablenote{Experiment, Ref.~\cite{nakamoto1986}}  &  & (1903)$^*$ & 
(0.157)\tablenote{Experiment, Ref.~\cite{hoy1975}} \\
on-top    & tilted      & 0.320     & 0.89 & 1.1704 & 2.3309 & --- & 0.353\\ 
bridge\tablenote{Metastable configuration}    &vertical & 0.148     & 0.94 & 1.1653 & 
2.8206 & --- & 0.313\\ 
fcc$^{c}$  &vertical & 0.122     & 1.14 & 1.1688 & 2.8576 & --- & 0.172 \\ \hline 
\end{tabular*}
\label{table:dft}
\end{table*}   

\begin{table*}[h]
\centering
\caption{Anderson model parameters for NO/Au(111)}
\begin{tabular*}{13cm}{@{\extracolsep{\fill}}l @{\SP}l@{}l@{\SP} l@{}l @{\SP} l@{}
l @{\SP} l @{\SP} l @{\SP} l}
site   & $\epsilon_e$ & $\epsilon_o$ & $\Gamma_e$ & $\Gamma_o$ & $U_e$ & $U_o$ & 
$U_{eo}$ & $J_H$ & $W_{eo}$\\ \hline
on-top   &   -0.92 & -1.02   & 0.33 & 0.068    & 2.24 & 2.04    & 1.67 & -0.807 & -0.134\\ 
bridge   &   -1.22 & -1.83   & 0.12 & 0.13      & 2.01 & 3.18    & 2.13 & -0.807 & -0.134\\ 
fcc        &   -1.00 & -1.00   & 0.22 & 0.22      & 2.45 & 2.45    & 1.83  & -1.06  & -0.178\\
\hline 
\end{tabular*}\\
{\scriptsize \textsf{All quantities in eV; SDs of $\epsilon_e$, $\epsilon_o$, $\Gamma_e$ and $\Gamma_o$ for the on-top site are 0.04, 0.01, 0.036, and 0.015, respectively.}}
\label{table:params}
\end{table*}

\end{document}